\documentclass{egpubl}
\usepackage{pg2026s}

\WsConferencePaper

\usepackage[T1]{fontenc}
\usepackage{dfadobe}  

\usepackage{cite}  
\BibtexOrBiblatex
\electronicVersion
\PrintedOrElectronic
\ifpdf \usepackage[pdftex]{graphicx} \pdfcompresslevel=9
\else \usepackage[dvips]{graphicx} \fi

\usepackage{egweblnk} 

\usepackage{amsmath,amsfonts}

\newcommand{\sysName}[0]{\textbf{LayoutShop}}

\usepackage{xspace}
\makeatletter
\DeclareRobustCommand\onedot{\futurelet\@let@token\@onedot}
\def\@onedot{\ifx\@let@token.\else.\null\fi\xspace}

\def\eg{{e.g}\onedot} 
\def\ie{{i.e}\onedot} 
 
 \def\vs{{vs}\onedot}
 
\def\etal{{et al}\onedot}
\makeatother

\title[\sysName]%
      {\sysName: Content-Constrained Exploratory Design of \\ Creative Article Layout}

\author[J. Li \& P. Xu]{
\parbox{\textwidth}{\centering 
    Jialuo Li\orcid{0009-0003-9964-6249} and
    Pengfei Xu\thanks{Corresponding author: Pengfei Xu (xupengfei.cg@gmail.com)}\orcid{0000-0003-4770-4374}
        }
        \\
{\parbox{\textwidth}{\centering CSSE, Shenzhen University, China}
}
}

\usepackage{xcolor}

\begin{document}

\teaser{
 \includegraphics[width=\linewidth]{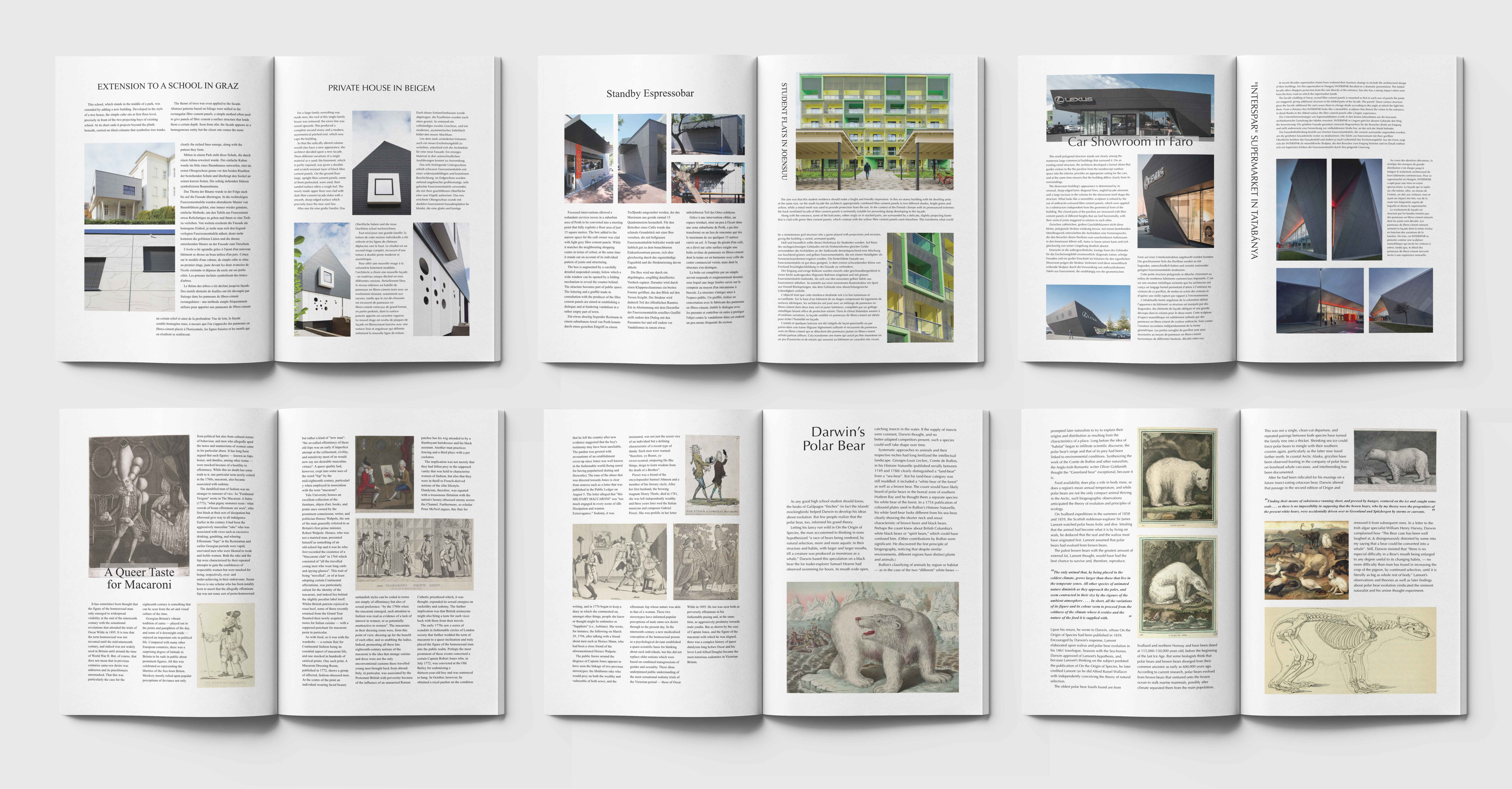}
 \centering
  \caption{The proposed \sysName\ can effectively help users create diverse, creative, and high-quality article layouts based on the contents of given articles. The first row displays single-page layouts (six articles) and the second row displays multi-page layouts (two articles).}
\label{fig:teaser}
}

\maketitle

\begin{abstract}
We present \sysName, a novel computational framework for designing creative layouts that frame a given article. Inspired by the actual article layout design process, we enable users to create or select layout templates for conceptualization. These templates help construct a layout design space to extract eligible layout structures. Our algorithm then determines the geometry of the extracted layout structures to frame the given article via an optimization approach. We then employ two neural networks for layout assessment, and the high-quality outputs are returned to users for selection. We conducted a user study to evaluate the framework's usability and the quality of the article layouts it produces. The results of the user study confirmed that our framework can effectively help users create high-quality article layouts. We will release the code of our framework upon the acceptance of this paper.

\printccsdesc   
\end{abstract}  

\section{Introduction} \label{sec:intro}

Proper arrangement of text and illustrations makes an article clear and visually appealing~\cite{lupton2014thinking, muller1996grid}. This involves choosing typefaces, setting leadings, and designing layouts. While typeface and leading are often standardized, the layout for text and illustrations can vary.
A simple way to create an article layout is to use a template, such as a \LaTeX\ or Microsoft Word template. These provide easy-to-use, conventional layouts.
However, creating more creative layouts, like a magazine page (Figure~\ref{fig:teaser}), requires significant creativity, skill, and effort, making the process tedious and challenging even for professionals.

To help users, many studies aim to automate layout generation. Some works~\cite{dayama2020grids, swearngin2020scout} automate GUI layouts by arranging known graphic elements, but these methods don't apply to article layouts where text boxes vary in number and shape.
Recently, learning-based methods~\cite{gupta2021layouttransformer, arroyo2021variational, zheng2019content, kikuchi2021constrained, jyothi2019layoutvae,jiang2022coarse, kong2022blt, chai2023layoutdm, inoue2023layoutdm, hui2023unifying} have made significant progress, producing diverse and creative layouts from training data.
However, these methods generate unstructured templates that contain sets of labeled boxes with fixed shapes and lack spatial relationships. As a result, they may not frame article content well or meet design requirements like typeface, leading, or page limits.
Thus, a gap remains between current layout-generation methods and fully automatic article layout design.

\begin{figure*}[t]
    \centering
    \includegraphics[width=\linewidth]{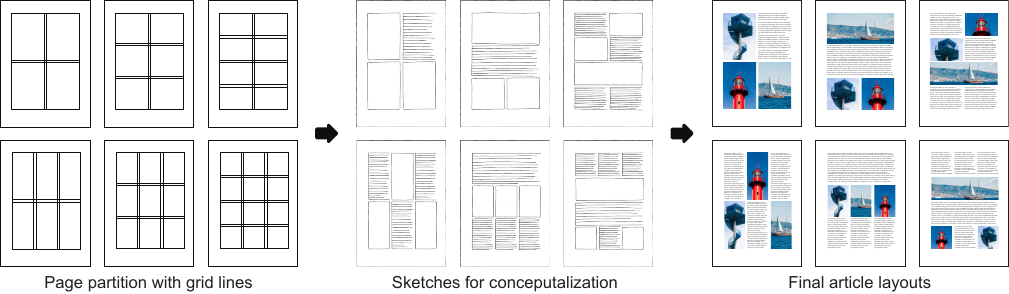}
    \caption{A typical procedure~\cite{muller1996grid} to create article layouts. With the aid of the grid system (left), a designer first creates small sketches to outline the desired article layouts (middle). Then the designer places an article's text and illustrations to create the final layouts (right).}
    \label{fig:procedure}
\end{figure*}

An article's layout is determined by its content. M{\"u}ller-Brockmann, a renowned designer, summarized the layout creation process in his book~\cite{muller1996grid}.
As shown in Figure~\ref{fig:procedure}, a designer partitions the page using a grid system, sketches the desired layout, creates a grid-based layout from the sketch, and refines it to fit the article's content~\cite{muller1996grid}. This trial-and-error process is often tedious and time-consuming. A computational framework could meaningfully assist users in designing article layouts.

We present \sysName, a computational framework for designing creative article layouts that frame a given article's content. While manually creating high-quality layouts is challenging, users can easily create or choose layout templates to express their preferences.
Although these templates may not perfectly frame an article’s content, they can define a design space when combined with layout blending techniques~\cite{xu2022hierarchical}.
This design space contains layouts that reflect user preferences and effectively frame content, though the optimal layouts are not immediately apparent.
To bridge layout exploration and content constraints, we introduce a computational workflow that couples a two-stage strategy (layout conceptualization and finalization) inspired by~\cite{muller1996grid} with dual neural assessment networks for layout quality filtering.
In layout conceptualization, we unify the created and selected layout templates into a compound structure~\cite{xu2022hierarchical} and extract potential layout structures that satisfy exact article requirements. In layout finalization, we resolve layout geometry via a novel article-specific mixed-integer quadratically constrained quadratic program (MIQCQP) formulation.
Layouts with feasible optimized geometries are retained as candidates.
To ensure high-quality results, we use two neural networks to assess layout quality from visual and structural perspectives. Only layouts approved by both networks are shown for user selection, with further refinement possible using editing tools or existing editors.

We demonstrate that our framework effectively enables users to create high-quality article layouts in various styles (see Figures~\ref{fig:teaser}, \ref{fig:user_input}, and \ref{fig:study1_examples}).
These layouts are based on real articles from \emph{Architecture+Detail}~\footnote{https://www.architecture-plus-detail.com/en/ad/} and \emph{The Public Domain Review}~\footnote{https://publicdomainreview.org}.
Layout templates for user selection are sourced from the Magazine dataset~\cite{zheng2019content} and processed to extract their structures~\cite{xu2022hierarchical}.
Generated article layouts conform to templates that users create or select.
We compare our method with GRIDS~\cite{dayama2020grids}, a method for creating regular layouts. The results confirm that our method is better suited for article layout creation.
We conducted a usability study with both novice users and experienced designers, asking them to create article layouts using our framework and traditional editors.
Statistics and user feedback confirmed the effectiveness of our framework in assisting article layout creation and also validated our layout assessment networks.
A follow-up study found that layouts created with our framework were superior to those manually created by both novice and experienced users using traditional editors.

\section{Related Work} \label{sec:related}

\paragraph*{Interactive layout design.} 
Interactive layout design methods help users efficiently create graphical layouts. 
Ambiguous intentions~\cite{Gross1996AmbiguousIA} and Sketchplore~\cite{Todi2016SketchploreSA} use interactive sketching for rapid design. 
Some approaches use heuristics to create visually pleasing layouts. O'Donovan \etal~\cite{ODonovan2014LearningLF} introduced an energy-based model for single-page graphic designs. GRIDS~\cite{dayama2020grids} generates GUI layouts using packing, alignment, and grouping.
Other systems offer design alternatives and feedback. DesignScape~\cite{o2015designscape} suggests both refined and novel layouts. Scout~\cite{swearngin2020scout} lets users rapidly explore alternatives and uses feedback to repair layouts. GUIComp~\cite{lee2020guicomp} provides real-time feedback on balance, alignment, and color unity. 
Recent methods~\cite{laine2021responsive, chen2023docdancer} help users create responsive layouts. 
Our approach generates article layouts from content and user-selected templates, and provides interactive tools for editing layouts to match user intent.

\paragraph*{Template-based layout creation.}
Template-based creation uses predefined patterns to automate layouts~\cite{Herring2009GettingIU}.
Some methods generate layouts by selecting from template libraries. 
Jacobs \etal~\cite{jacobs2003adaptive} used dynamic programming and constraints for template selection. Schrier \etal~\cite{schrier2008adaptive} extended this for dynamic documents. Yang \etal~\cite{yang2016automatic} introduced topic-specific templates.
Other research adapts and transfers existing designs. Dayama \etal~\cite{Dayama2021InteractiveLT} transfer content to templates; Bricolage~\cite{kumar2011bricolage} retargets web layouts using reusable structures. Kikuchi \etal~\cite{Kikuchi2021ModelingVC} optimizes webpages with visual containment.
Some methods combine template retrieval and data-driven recombination. Damera \etal~\cite{DameraVenkata2011ProbabilisticDM} learn spatial relationships from templates. Faceoff~\cite{zheng2019faceoff} recommends style rules using structure-aware retrieval.
Our method uses templates to generate layouts that fit article content.

\paragraph*{Optimization-based methods.} 
These methods formulate design principles and optimize layouts accordingly~\cite{xu2019global, dayama2020grids, Jiang2020ORCSolverAE, Kikuchi2021ModelingVC}. Jacobs \etal~\cite{jacobs2003adaptive} adapted articles to different page sizes using layout templates, highlighting the need for proper templates.
O'Donovan \etal~\cite{ODonovan2014LearningLF} used an energy-based model for features like alignment and saliency. Jiang \etal~\cite{Jiang2020ORCSolverAE} proposed a solver for OR-constrained UI layouts. Shiripour \etal~\cite{shiripour2021grid} introduced an evolutionary algorithm for non-overlapping layouts. 
However, these methods may not satisfy individual user preferences.
In contrast, our method addresses user preferences and generates novel article layouts.

Recent studies use deep neural networks to score and optimize layouts~\cite{tabata2019automatic, Duan2020OptimizingUI}. Probabilistic models estimate designer preferences~\cite{DameraVenkata2011ProbabilisticDM}, while other networks predict scores for optimization~\cite{zhao2018characterizes, Duan2020OptimizingUI}. Kikuchi \etal~\cite{kikuchi2021constrained} optimize layouts in latent space under user constraints. Our method blends heuristic rules with data-driven assessments, optimizing for alignment, spacing, and layout principles using assessment networks.

\paragraph*{Learning-based generative methods.} 
Learning-based generative methods use large datasets~\cite{zheng2019content, Deka2017RicoAM, Zhong2019PubLayNetLD} to generate diverse layouts. Early methods~\cite{Wang2018DeepCP, Ritchie2018FastAF, Wu2019DatadrivenIP} used convolutional networks for raster images, but most now operate on vectorized layouts. NDN~\cite{Lee2019NeuralDN} and LayoutGAN~\cite{Li2019LayoutGANGG} use graphs to model relations, but require fixed element counts. Hierarchical approaches~\cite{Li2018GRAINS, Patil2019READRA, jiang2022coarse} predict layouts in a top-down manner.
Other methods treat layouts as sequences. Autoregressive decoders with attention~\cite{gupta2021layouttransformer, arroyo2021variational, Wang2020SceneFormerIS, xu2026gtlayout, Yang2025OrderMatters} or set-based approaches~\cite{Paschalidou2021ATISSAT} predict layout elements. Diffusion models~\cite{chai2023layoutdm, inoue2023layoutdm, he2023diffusion, levi2023dlt, chen2024towards, chai2023two, jiang2025illustration, guo2025contentdm} now achieve state-of-the-art results.

Controllable layout generation uses bidirectional Transformers for layout completion~\cite{kong2022blt, hui2023unifying}.
Jiang \etal~\cite{Jiang2022LayoutFormerCG} restricts decoding space to avoid undesired layouts.
However, these networks may not guarantee layout quality.
Lin \etal~\cite{lin2023parse} use language models and Transformers for text-to-layout generation, lowering design barriers.
LayoutPrompter~\cite{lin2023layoutprompter} uses large language models for constraint-aware and text-to-layout generation.
StructLayoutFormer~\cite{hu2025structlayoutformer} uses Transformers with structure serialization to generate 2D layouts and hierarchies.

Large language models (LLMs) and multi-modal LLMs now drive layout generation, leveraging implicit design knowledge and spatial reasoning~\cite{tang2024layoutkag, zhang2024vascar, zhang2025smaller, cheng2025graphic, lu2025uni}.
These models capture high-level relations but miss hard constraints like alignment.
Generated layouts may not match specified content, but our method explicitly considers article content.


\paragraph*{Content-aware layout generation.} 
Content-aware layout generation considers all content (text, images, etc.) during design. 
Neural networks arrange poster or banner layouts by detecting salient regions~\cite{zhang2020smarttext, vaddamanu2022harmonized, shabani2024visual}. 
Qiang \etal~\cite{qiang2016learning} generates hierarchical poster layouts by recursively partitioning space. 
Vinci~\cite{guo2021vinci} lets designers input images and attributes for conditional generation.
FlexDM~\cite{inoue2023towards} and MarkupDM~\cite{Kikuchi2025MarkupDM} use unified multi-modal models for layout generation and content filling.
Recent work leverages large language models and structured reasoning for adaptable, coherent layout generation~\cite{Hsu2025PosterO, Teng2025PosterCoT, Chen2025T-Stars-Poster}.
However, these methods overlook user preferences and creativity.

Both semantic content and layout principles must be respected. 
Zhou \etal~\cite{Zhou2022CompositionawareGL} use cross-attention in GANs for content-layout relationships. 
Cao \etal~\cite{Cao2022GeometryAV} add geometry-aligned fusion for image-conditioned layouts. 
Hsu \etal~\cite{Hsu2023PosterLayoutAN} convert layouts into sequences for GAN learning. 
LayoutDETR~\cite{yu2024layoutdetr} treats layout generation as detection, predicting element positions and relations.
PosterLlama~\cite{seol2024posterllama} uses HTML and language models for rich, coherent layouts. 
RALF~\cite{horita2024retrieval} retrieves similar examples to guide autoregressive layout generation.
Beyond single-frame graphic layouts, sequential page layout generation, such as manga and comic paneling, has also been studied to handle storytelling constraints and reading flow~\cite{cao2012automatic, yang2021automatic, chen2024manga}.
Most existing methods focus on sparse graphic layouts (posters, banners) or panel-based narrative layouts, whereas our approach targets complex, multi-page article layouts that demand strict typographical consistency, multi-column fitting, and continuous text-illustration balance.



\begin{figure*}[t]
    \centering
        \includegraphics[width=\linewidth]{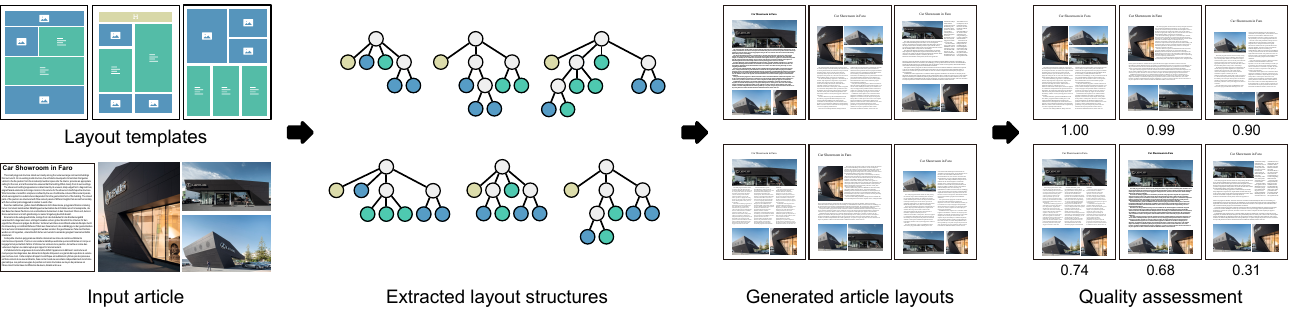}
    \caption{The pipeline of our framework. The input to our framework consists of an article and several layout templates (Col. 1). Our algorithm extracts a set of layout structures that may potentially frame the input article's content (Col. 2). Our algorithm optimizes the geometry of the extracted layout structures to frame the input article's content (Col. 3). Two neural networks assess the qualities of the generated article layouts (Col. 4).}
    \label{fig:pipeline}
\end{figure*}

\section{Article Layout Design Procedure}\label{sec:procedure}

Modern typography and article layout methods have evolved over the past century~\cite{muller1996grid}. The grid system~\cite{muller1996grid, lupton2014thinking} is a key advancement, using grid lines (Figure~\ref{fig:procedure}, left) to guide graphic placement. This system helps designers partition pages and organize content into well-aligned, regular layouts.

When designing an article layout, the designer creates a grid based on the article's content, following a process outlined in~\cite{muller1996grid}.
First, the designer decides on format details, including typeface, leading, page limits, etc. Then the designer sketches possible layouts (Figure~\ref{fig:procedure}, middle).
These sketches reflect the article's content and may show different layout structures.
For example, layouts may vary in column number and illustration placement.
The designer compares the sketches, discards unsuitable ones, and applies a grid to the chosen design, placing text and illustrations accordingly. Usually, this initial layout doesn't fit the content perfectly.
As~\cite{muller1996grid} notes, ``\emph{It is rare to find the final solution the first time.}'' Designers must repeatedly adjust the layout geometry to achieve a satisfactory result (Figure~\ref{fig:procedure}, right).

This design process inspired our algorithm for automatic article layout generation, which mirrors the design process's two stages: layout conceptualization and finalization. In the conceptualization stage, users sketch potential layouts to frame the article's content.
This creative task is time-consuming, even for professionals.
To this end, we use the method from~\cite{xu2022hierarchical} to define a layout design space, thereby enabling the automatic extraction of eligible structures (Section~\ref{sec:structure_extraction}).
In the finalization stage, users place content and adjust the layout's geometry to achieve a high-quality result.
We automate this adjustment using a mixed-integer quadratically constrained quadratic program (MIQCQP), generating layouts by solving this optimization problem (Section~\ref{sec:optimizaion}).

\begin{figure*}[t]
    \centering
    \includegraphics[width=0.95\linewidth]{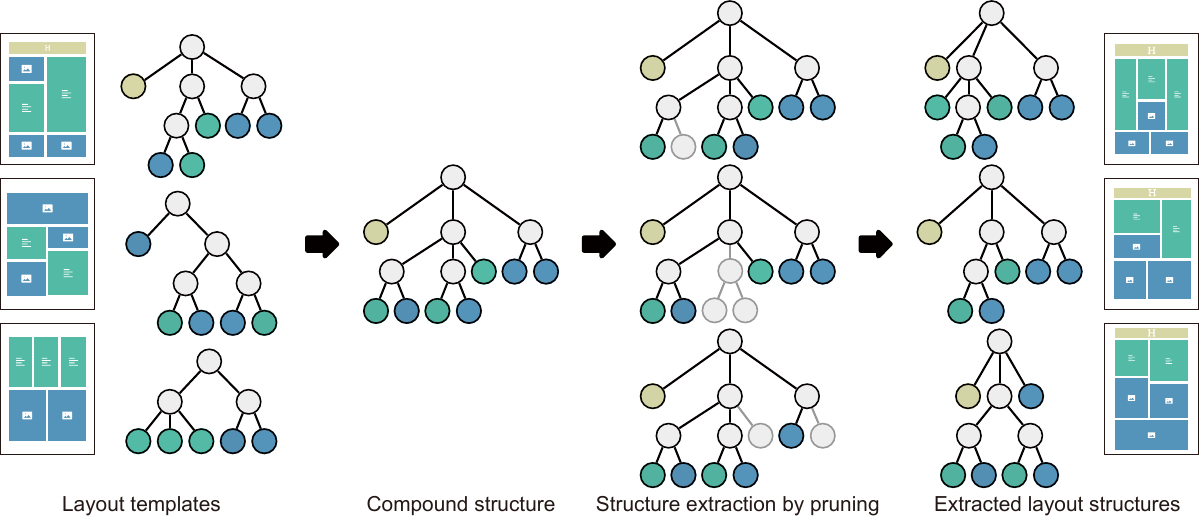}
    \caption{Layout conceptualization by structure extraction. The input of our approach is layout templates selected or created by users (Col. 1). The input layout templates help construct a compound tree (Col. 2). By pruning the leaves in the compound tree (Col. 3), different novel layout structures can be obtained (Col. 4).}
    \label{fig:structure_extraction}
\end{figure*}

\section{Content-Constrained Article Layout Generation} \label{sec:method}

Figure~\ref{fig:pipeline} illustrates our framework's pipeline. The input is an article and user-selected or created layout templates (Figure~\ref{fig:pipeline}, Col. 1). In the conceptualization stage (Section~\ref{sec:structure_extraction}), our algorithm extracts eligible layout structures to frame the article (Figure~\ref{fig:pipeline}, Col. 2). In the finalization stage (Section~\ref{sec:optimizaion}), it optimizes the geometry of the extracted layout structures to frame the input article's content (Figure~\ref{fig:pipeline}, Col. 3). After obtaining the article layouts, we exploit two neural networks to assess the qualities of the generated layouts (Figure~\ref{fig:pipeline}, Col. 4, Section~\ref{sec:networks}).

\subsection{Layout conceptualization by structure extraction} \label{sec:structure_extraction}

As discussed in Section~\ref{sec:procedure}, designers typically sketch several possible layouts to match their preferences and frame the article’s content. This process is time-consuming.
Alternatively, users can select existing layout templates, but these templates may not fit the article's content.
Testing every template for compatibility is impractical.
This search may also fail to satisfy user preferences.

A better approach is to automatically generate new layouts based on user preferences using the layout blending method from~\cite{xu2022hierarchical}. This technique constructs a design space from a small set of structured layouts, producing new layouts similar to the inputs and suitable for geometry optimization. We represent each layout as a tree: leaf nodes are graphic elements (headline, text, or illustration), and branch nodes represent the arrangements of their children in the same order as in the actual layout, making each tree unique (see Figure~\ref{fig:structure_extraction}).
Notably, selected templates do not need to match the content in the number of element types; they just need to combine into a compound structure with enough elements. For example, Figure~\ref{fig:pipeline} shows that two of the three templates lack headline elements, yet the content includes a headline.
We then devise the following approach to generate the layout structure.

\paragraph*{Single-page layout structure.}
We start by describing the case where the article layout is limited to a single page.
Our approach uses three user-selected or created layout templates (Figure~\ref{fig:structure_extraction}, Col. 1). While current templates come from the \emph{Magazine} dataset~\cite{zheng2019content}, integrating more external sources and custom templates (Figure~\ref{fig:user_input}) further expands the template pool to ensure generation diversity and quality.
Template structures are represented as trees, estimated using the method in~\cite{xu2022hierarchical}.
We unify the input templates into a compound tree (Figure~\ref{fig:structure_extraction}, Col. 2) as in~\cite{xu2022hierarchical}, then extract eligible substructures to frame the article. An eligible structure must match the article's number of headlines and illustrations, but the number of text elements can vary, as text can be split without affecting readability. We generate eligible structures by pruning leaves in the compound tree until requirements are met (Figure~\ref{fig:structure_extraction}, Col. 3); if a branch has no children left, it is pruned too.
To retain a user-created template in the final layout, we prevent its leaves from being pruned (see Figure~\ref{fig:user_input}).
Different pruning sequences yield distinct layout structures (Figure~\ref{fig:structure_extraction}, Col. 4), making this a combinatorial problem with ample options.

\paragraph*{Multi-page layout structure.}
A multi-page article layout is represented as a forest, with each tree corresponding to a single page. We extract the multi-page layout structure from a compound forest of multiple trees by pruning leaves until the structure meets the same requirements as for single-page layouts. Optionally, each page’s substructure can be required to have the same number of columns. This approach yields ample layout structures.

\paragraph*{Eligible structure filtering.}
Note that extracted structures only meet basic eligibility; some may still not frame the article's content well.
Using theoretical design rules to discard unqualified layouts is cumbersome.
Only eligible structures yield feasible solutions during geometry optimization. Thus, we focus on collecting enough structures now and filter them in the optimization stage, mirroring the article layout design process in Section~\ref{sec:procedure}.

\subsection{Layout finalization by geometry optimization}\label{sec:optimizaion}

\begin{figure}[t]
    \centering
    \includegraphics[width=\linewidth]{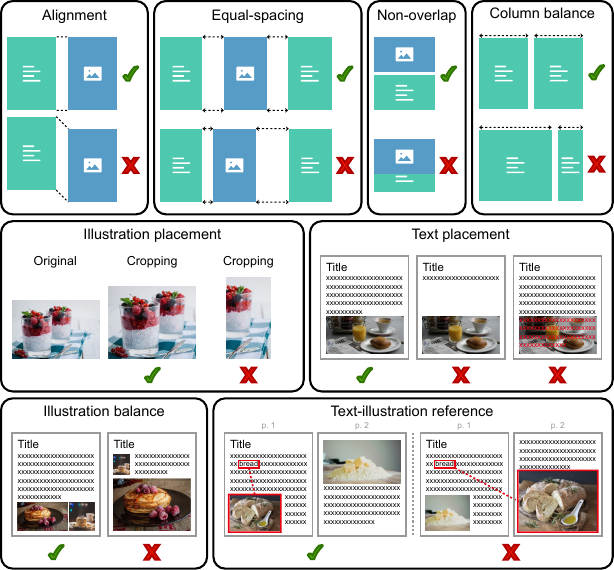}
    \caption{The illustration of the criteria for a good article layout.}
    \label{fig:criteria}
\end{figure}

We now have layout structures that can potentially frame the article’s content. Instead of manually fine-tuning the geometry, as discussed in Section~\ref{sec:procedure}, we formulate the requirements for a satisfactory layout as an optimization problem with structure constraints. Drawing on established design principles~\cite{muller1996grid, lupton2014thinking, dayama2020grids}, we define the followig key criteria for a good article layout (see Figure~\ref{fig:criteria}):

\begin{itemize}
\item 
\textit{Alignment.} 
Elements should be well-aligned.

\item
\textit{Equal-spacing.} 
Spacing between elements should be uniform.

\item 
\textit{Non-overlap.} 
Elements should not overlap.

\item
\textit{Text placement.} 
Text should not overflow or leave excessive blank space.

\item
\textit{Illustration placement.} 
Illustrations should be properly framed without excessive cropping.

\item 
\textit{Text-illustration reference.}
Illustrations should be placed near relevant text.

\item
\textit{Column balance.} 
Columns should have similar widths for readability.

\item
\textit{Illustration balance.} 
Illustrations should be similar in size for consistency.
\end{itemize}

Some of the above criteria, \ie, alignment, equal-spacing, non-overlap, text placement, illustration placement, and Text-illustration reference, are mandatory for a good layout. The remaining criteria, \ie, column balance, illustration balance, and user specification, are preferred but not always attainable. These categories define our optimization objectives and constraints.

\paragraph*{Mathematical formulation.}
Given a layout tree, we must determine its geometry to properly frame the article's content.
We formulate this as a mixed-integer quadratically constrained quadratic program (MIQCQP).
We introduce the following notations. Let $\mathcal{V}$ be the vertex set of the layout tree, divided into the leaf set $\mathcal{V}_{\mathrm{leaf}}$ and branch set $\mathcal{V}_{\mathrm{bran}}$. The leaf set $\mathcal{V}_{\mathrm{leaf}}$ is further split by semantic label into the headline set $\mathcal{V}_{\mathrm{head}}$, text set $\mathcal{V}_{\mathrm{text}}$, and illustration set $\mathcal{V}_{\mathrm{ills}}$.
The optimization variables are classified into four sets:

\begin{itemize}
\item
\textit{Element geometry.} 
The first set of variables specifies the geometry of layout elements.
For each vertex $v_i\in\mathcal{V}$ (leaf or branch), its position and size $\{x_i, y_i, w_i, h_i\}$ are variables.
The coordinate origin is the top-left corner, with $x$ to the right and $y$ downward.

\item
\textit{Margins.} 
The second set, $\{m_{\mathrm{l}}, m_{\mathrm{r}}, m_{\mathrm{t}}, m_{\mathrm{b}}\}$, represents the distances from the type area to page boundaries.

\item
\textit{Spacings.} 
The third set, $\{s_{\mathbf{h}}, s_{\mathbf{v}}\}$, are the horizontal and vertical spacings between neighboring elements.

\item
\textit{Illustration correspondence indicators.}
The fourth set consists of correspondence indicators between layout illustration elements and the article's illustrations.
Let $\mathcal{N}$ be the set of article illustrations.
For each illustration leaf $v_i\in\mathcal{V}_{\mathrm{ills}}$ and illustration $n_j\in\mathcal{N}$, the binary indicator $b_{i,j}$ denotes whether $v_i$ frames $n_j$.

\end{itemize}

We denote the complete variable set as $\mathcal{Z}$: $\mathcal{Z} = \{x_i, y_i, w_i, h_i|v_i\in\mathcal{V}\} \cup\{m_{\mathrm{l}}, m_{\mathrm{r}}, m_{\mathrm{t}}, m_{\mathrm{b}}\}\cup\{s_{\mathbf{h}}, s_{\mathbf{v}}\}\cup \{ b_{i,j}|v_i\in\mathcal{V}_{\mathrm{ills}}, n_j\in\mathcal{N}\}$.

\paragraph*{Objective function.}
The objective function, based on column balance, illustration balance, and user specification, is:
\begin{equation}
    E= \alpha E_{\mathrm{coln}} + \beta E_{\mathrm{ills}}+ \gamma E_{\mathrm{user}},
\end{equation}
with
\begin{equation}\label{eqn:obj}
\begin{aligned}
    &E_{\mathrm{coln}}= \sum_{v_i, v_j\in\mathcal{V}_{\mathrm{text}}}^{v_i \neq v_j }(w_i-w_j)^2 , \\
    &E_{\mathrm{ills}}= \sum_{v_i, v_j\in\mathcal{V}_{\mathrm{ills}}}^{v_i \neq v_j } ((w_i-w_j)^2 + (h_i-h_j)^2), \\
    &E_{\mathrm{user}}= \sum_{z\in\mathcal{Z}} \delta(z)(z - z')^2.\\
\end{aligned}
\end{equation}
Here, $\alpha=1$, $\beta=1$, $\gamma=10000$ control each term's weight, with a large $\gamma$ to emphasize user editing. These objective weights were determined and refined via iterative empirical tuning on a validation subset of article layouts. $\delta(z)$ is an indicator function equal to $1$ if the individual variable $z \in \mathcal{Z}$ is updated to a preferred value $z'$ by the user during interactive editing (Section~\ref{sec:ui}), or $0$ otherwise.

\paragraph*{Constraints.}

Constraints are based on alignment, equal-spacing, non-overlap, text placement, illustration placement, and text-illustration reference. The layout structure facilitates easy formulation. Common constraints, such as alignment~\cite{dayama2020grids} and equal-spacing~\cite{xu2019global}, are reiterated here for clarity.

For alignment, if vertices $v_i$ and $v_j$ share a parent, they are aligned. Vertical and horizontal alignments are defined as:
\begin{subequations}
\begin{align}
    x_i &= x_j, \quad \mathrm{and} \quad x_i+w_i = x_j+w_j; \\
    y_i &= y_j, \quad \mathrm{and} \quad y_i+h_i = y_j+h_j.
\end{align}
\end{subequations}

The equal-spacing constraint uses spacing variables. For neighboring siblings $v_i$ and $v_j$ in the layout tree, the horizontal and vertical equal-spacing constraints are defined as:
\begin{subequations}
\begin{align}
    x_i+w_i+s_{\mathrm{h}} &= x_j, \quad \mathrm{with} \quad s_{\mathrm{h}}^{\mathrm{min}} \leq s_{\mathrm{h}} \leq s_{\mathrm{h}}^{\mathrm{max}}; \\
    y_i+h_i+s_{\mathrm{v}} &= y_j, \quad \mathrm{with} \quad s_{\mathrm{v}}^{\mathrm{min}} \leq s_{\mathrm{v}} \leq s_{\mathrm{v}}^{\mathrm{max}}.
\end{align}
\end{subequations}
Here, $s_{\mathrm{h}}^{\mathrm{min}}$, $s_{\mathrm{h}}^{\mathrm{max}}$, $s_{\mathrm{v}}^{\mathrm{min}}$, and $s_{\mathrm{v}}^{\mathrm{max}}$ define the user-specified spacing ranges.

Horizontal and vertical equal-spacing constraints prevent sibling overlap. To fully ensure non-overlap, the parent vertex's geometry must tightly enclose its children. For a branch vertex $v_i$ with leftmost/topmost child $v_j$ and rightmost/bottommost child $v_k$, the horizontal and vertical non-overlap constraints are:
\begin{subequations}
\begin{align}
    &x_i=x_j, \quad x_i+w_i= x_k+w_k, \quad y_i = y_j = y_k, \quad h_i = h_j = h_k; \\
    &y_i=y_j, \quad y_i+h_i= y_k+h_k, \quad x_i = x_j = x_k, \quad w_i = w_j = w_k.
\end{align}
\end{subequations}

The page size constrains the article layout geometry. For root vertex $v_{0}$, the constraint is:
\begin{equation}
\begin{aligned}
    &x_0 = m_{\mathrm{l}}, &\quad
    &y_0 = m_{\mathrm{t}}, &\\
    &x_0+w_0+ m_{\mathrm{r}}= W, & \quad
    &y_0+h_0+ m_{\mathrm{b}}= H, &\\
\end{aligned} 
\end{equation}
with
\begin{equation}
\begin{aligned}
    &m_{*}^{\mathrm{min}} \leq m_{*} \leq m_{*}^{\mathrm{max}},  &\forall *\in\{\mathrm{l}, \mathrm{r}, \mathrm{t},\mathrm{b}\}, \\
\end{aligned}
\end{equation}
where $W$ and $H$ are the width and height of the page, and are specified by users for designing different article pages. $m_{*}^{\mathrm{min}}$ and $m_{*}^{\mathrm{max}}$, $*\in\{\mathrm{l}, \mathrm{r}, \mathrm{t},\mathrm{b}\}$, confine the ranges of the margins. They can be specified by users for different design purposes. 

Text placement requires the total area of text elements to be reasonable. Given the typeface, font size, and line leading, we estimate the character dimensions and text area ($A_{\mathrm{text}}$ and $A_{\mathrm{head}}$) via the Qt library API to set the allowable range. The constraint is:
\begin{equation}
\begin{aligned}
    &A_{\mathrm{text}}^{\mathrm{min}} \leq \sum_{v_i\in\mathcal{V}_{\mathrm{text}}} w_ih_i \leq A_{\mathrm{text}}^{\mathrm{max}}, \\
    &A_{\mathrm{head}}^{\mathrm{min}} \leq \sum_{v_i\in\mathcal{V}_{\mathrm{head}}} w_ih_i \leq A_{\mathrm{head}}^{\mathrm{max}}, \\
\end{aligned}
\end{equation}
where $A_{\mathrm{text}}^{\mathrm{min}}$, $A_{\mathrm{text}}^{\mathrm{max}}$, $A_{\mathrm{head}}^{\mathrm{min}}$, and $A_{\mathrm{head}}^{\mathrm{max}}$ are the area ranges for text and headline elements. 
They are determined by the occupation area of the article's text ($A_{\mathrm{text}}$) and headline ($A_{\mathrm{head}}$). 
Users may also specify them for particular design needs.

Illustration placement requires that aspect ratios do not change substantially. Since correspondence between illustrations and layout elements is unknown, we define the following constraint using binary variables:
\begin{equation}\label{eqn:page_size}
\begin{aligned}
&r_j^{\mathrm{min}} \leq \dfrac{w_i}{h_i} \leq r_j^{\mathrm{max}}, \quad \mathrm{if}\; b_{i, j} = 1, \\
&\forall v_i \in \mathcal{V}_{\mathrm{ills}},\; n_j \in \mathcal{N}, 
\end{aligned}
\end{equation}
and
\begin{equation}
\begin{aligned}
    \sum_{v_i\in\mathcal{V}_{\mathrm{ills}}}b_{i,j}=1, \quad &\forall n_j\in\mathcal{N},\\
    \sum_{n_j\in\mathcal{N}}b_{i,j}=1, \quad &\forall v_i\in\mathcal{V}_{\mathrm{ills}},\\
\end{aligned}
\end{equation}
where $r_j^{\mathrm{min}}$ and $r_j^{\mathrm{max}}$ define the range of $n_j$'s aspect ratio. They are based on $n_j$'s original aspect ratio $r_j$, and we set $r_j^{\mathrm{min}} = 0.8r_j$ and $r_j^{\mathrm{max}} = 1.25r_j$ in our implementation.

Text-illustration reference requires placing an illustration on the same page as its reference text and is applicable only to multi-page layouts. Given the document typography, we estimate the reference text's exact line position using the Qt library API to determine its residing page, then constrain the illustration correspondence indicators accordingly. If a layout structure cannot satisfy this, we discard it and try another.

We use the Gurobi solver~\cite{gurobi} for the MIQCQP.
Since the number of variables is small, this optimization problem can be solved efficiently: if a feasible solution exists, it can be obtained in less than 30 milliseconds. 
However, a feasible solution may not always exist if a layout structure cannot meet all constraints.
If computation exceeds a threshold (100 milliseconds in our implementation), we terminate and try another layout structure. Over $50\%$ of layout structures extracted during layout conceptualization are usually valid, so our framework typically generates a dozen layouts per second, making it practical for use.

\subsection{Data-driven layout assessment}\label{sec:networks}

Our framework typically generates high-quality layouts by incorporating key requirements into the objective function and constraints. However, certain high-level design qualities—such as visual hierarchy, semantic emphasis, and overall aesthetics—are difficult to formalize mathematically as explicit optimization constraints. Since our framework efficiently produces many layouts, we use a data-driven approach to evaluate these high-level qualities, filtering out unappealing candidate layouts for user selection.
Layout quality assessment should consider visual appearance, which captures the overall impression, and structural rationality, reflecting functional usage.
We use two neural networks, \ie, the visual and structural assessment networks, to evaluate layout quality.

\paragraph*{Visual assessment network.}
We use an image convolutional neural network, similar to the scoring network in~\cite{zhao2018characterizes}, to assess the visual appearance of the layout. Training uses positive and negative layout pairs $(l^+, l^-)$. Each layout $l$ is rendered as an image $I_l$ and scaled to $216\times296$.
To eliminate illustration texture effects, rendered layouts use color regions for semantic elements. During training, sample pairs $(I_{l^+}, I_{l^-})$ are input to the network, with the following loss function:
\begin{equation}\label{eqn:visual_loss}
\begin{aligned}
    L_{\mathrm{v}} = \max(0, m-S_{\mathrm{v}}(I_{l^+})+S_{\mathrm{v}}(I_{l^-})).
\end{aligned}
\end{equation}
Here, $S_{\mathrm{v}}(I_{l})$ is the network’s output indicating visual quality in $[0, 1]$, and $m=0.5$ is the margin between positive and negative samples.
To train this network, we adopt the Adam optimizer with an initial learning rate of $0.001$, a weight decay of $0.001$, a dropout rate of $0.5$, and a batch size of $32$.

\paragraph*{Structural assessment network.}
We use a graph convolutional neural network to assess structural rationality. Since layout structures are represented as trees, we use an architecture~\cite{tong2020directed} for directed graphs in our network. Training this network also uses positive and negative layout pairs. For each pair $(l^+, l^-)$, each layout $l$ is converted to a directed graph $G_l$ based on its tree structure: vertices become nodes storing geometry and type, with directed edges from children to parents.
During training, sample pairs $(G_{l^+}, G_{l^-})$ are input to the network. After convolutions, the root node's feature passes through a fully connected layer to produce the output. The loss function matches Equation~\ref{eqn:visual_loss}:
\begin{equation}\label{eqn:structure_loss}
\begin{aligned}
    L_{\mathrm{s}} = \max(0, m-S_{\mathrm{s}}(G_{l^+})+S_{\mathrm{s}}(G_{l^-})).
\end{aligned}
\end{equation}
Here, $S_{\mathrm{s}}(G_{l})$ is the network output in $[0, 1]$. Training settings match those of the visual assessment network.

\begin{figure}[t]
    \centering
    \includegraphics[width=\linewidth]{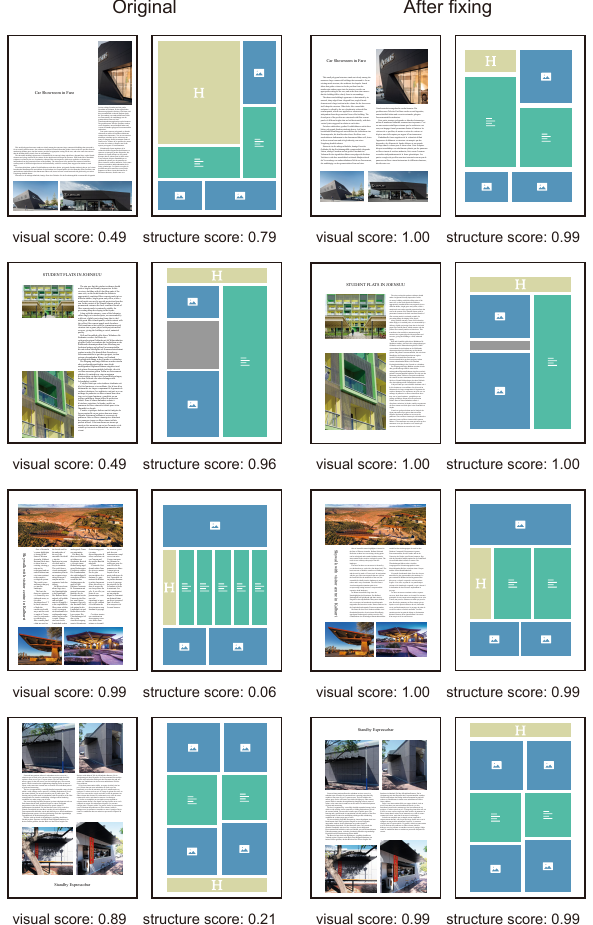}
    \caption{These examples clearly illustrate that the visual assessment network is sensitive to visual appearance, and the structural assessment network focuses on structural rationality. Left: Low-quality article layouts that have visual or structural artifacts. Right: The article layouts after fixing the artifacts.}
    \label{fig:network_example}
\end{figure}

\paragraph*{Sample pair preparation.}
Both networks require positive and negative layout pairs for training. Positive samples come from the Magazine~\cite{zheng2019content} and PubLayNet~\cite{Zhong2019PubLayNetLD} datasets, filtered to $3,307$ layouts with enough elements; structures are estimated as in~\cite{xu2022hierarchical}. We focus on distinguishing acceptable from clearly inadequate layouts, so negative samples are poor-quality layouts generated for each positive sample by recursively partitioning the page with random splits and label assignments, matching element count and labels. This yields $3,307$ negative samples. Each positive sample is paired with five negative samples, for a total of $16,535$ pairs. The training/testing split is $9:1$ for both networks.

\paragraph*{Assessment with two networks.}
We train the networks separately and use them jointly for layout assessment. Figure~\ref{fig:network_example} shows examples: the visual assessment network focuses on appearance, while the structural one emphasizes rationality.
To combine both networks, we score layout $l$ as $S(l) = \min(S_{\mathrm{v}}(l), S_{\mathrm{s}}(l))$, returning layouts with scores above $0.5$ as a lightweight binary filter to prune clearly inadequate candidates. Since the networks are trained via pairwise ranking loss, the output score represents a relative acceptability score indicating structural and visual alignment with high-quality designs, rather than a calibrated probability. Thus, a layout's score can increase significantly with even a single improvement.

Recent advances in vision-language models could allow the use of foundation models for assessment, but we use these two networks for the following reasons.
First, these networks explicitly capture the distinct visual and structural roles, making assessment interpretable.
Second, training with our data avoids domain mismatch: foundation models are tuned for natural images, not structured layouts. In addition, these lightweight networks are more efficient to train.

\section{User Interface} \label{sec:ui}

\begin{figure}[!t]
    \centering
    \includegraphics[width=\linewidth]{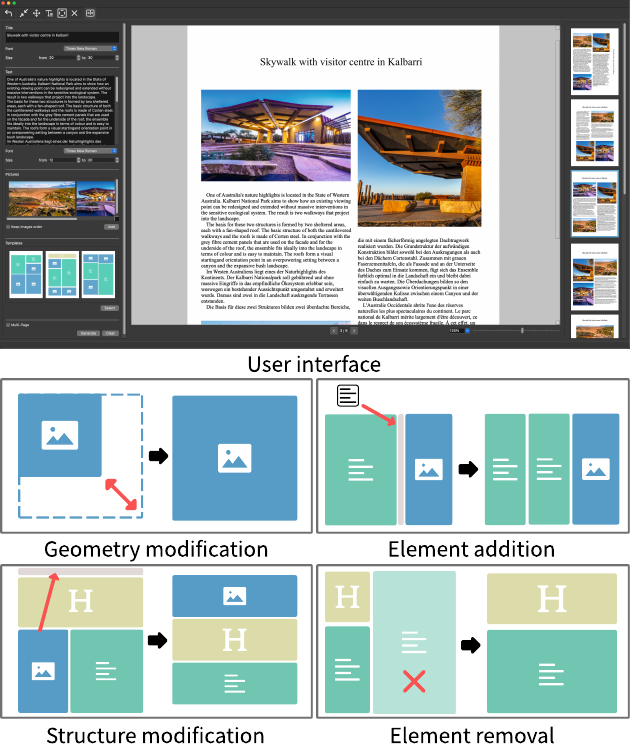}
    \caption{Top: the interface of our framework. Bottom: the illustrations of the interactive tools we provided.}
    \label{fig:ui}
\end{figure}

Figure~\ref{fig:ui} shows the interface of our framework. 
Users enter article content via widgets in the left panel or load it from a JSON file. Next, users select at least three layout templates using a widget or create their own; templates use abstract icons for layout elements. Clicking the generation button automatically creates article layouts, which appear in the right panel within seconds, ranked by our assessment networks. Layouts display as a list or grid. Users can browse, select, and view a layout in the main panel, then edit it with interactive tools. Final layouts can be saved in multiple formats, including SVG, enabling users to export generated results into professional vector editors for precise manual fine-tuning without re-optimization.

\begin{figure}[t]
 
    \centering
    \includegraphics[width=\linewidth]{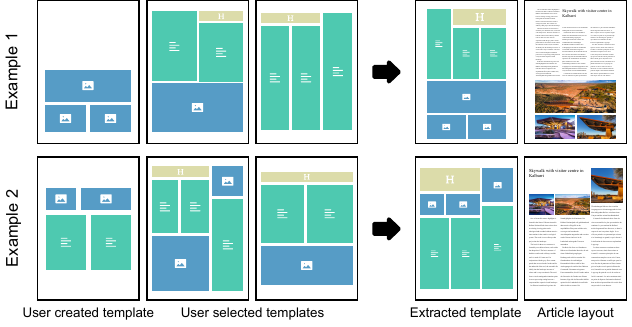}
    \caption{Our framework allows users to create a partial layout template to specify their design purposes. Combining other layout templates can obtain a complete article layout that satisfies their requirements.}
    \label{fig:user_input}
\end{figure}

\begin{figure}[t]
    \centering
    \includegraphics[width=\linewidth]{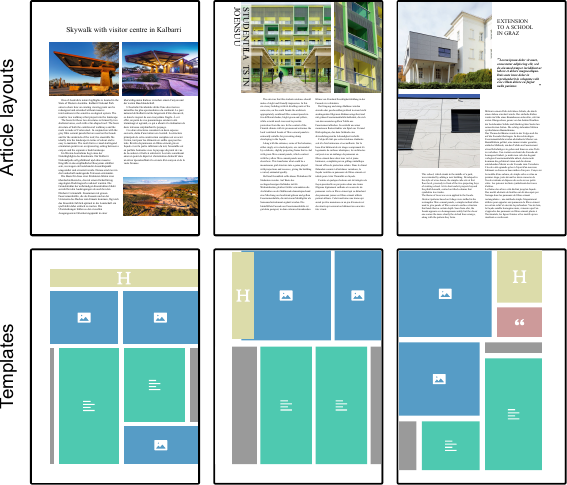}
    \caption{Our framework can help produce article layouts with special designs.}
    \label{fig:special}
\end{figure}

\paragraph*{Geometry modification tool.} 
This tool lets users modify layout geometry (Figure~\ref{fig:ui}, Row 2, left) by dragging element boundaries. Adjusted geometry is recorded for optimization (see Equation~\ref{eqn:obj}).
    
\paragraph*{Structure modification tool.} 
This tool allows users to modify layout structure (Figure~\ref{fig:ui}, Row 3, left) by dragging elements. The element's position determines which substructure it belongs to; after release, the structure updates, and geometry is recomputed.

\paragraph*{Element addition tool.} 
This tool lets users insert new elements into the layout (Figure~\ref{fig:ui}, Row 2, right) by dragging to a desired location. After release, the element is inserted, and the geometry is recomputed. To avoid mismatches, only text or padding elements can be added, as headlines and illustrations are fixed.

\paragraph*{Element removal tool.} 
This tool allows users to remove elements from the layout (Figure~\ref{fig:ui}, Row 3, right) by selecting and deleting them. Only text and padding elements can be removed to prevent mismatches.

\begin{figure}[t]
    \centering
    \includegraphics[width=\linewidth]{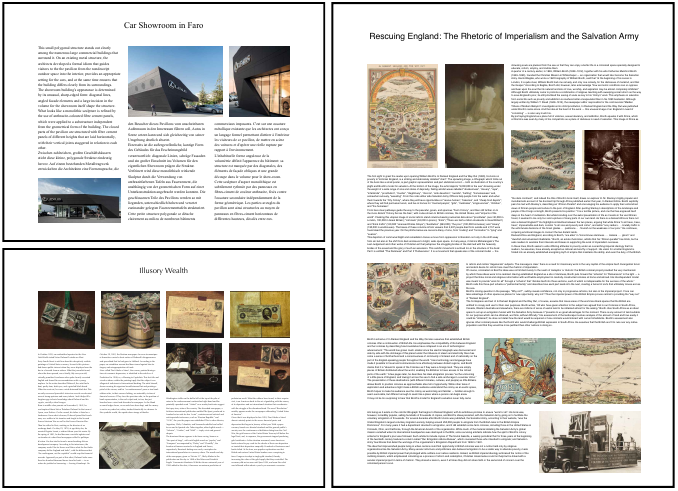}
    \caption{Our framework can generate high-quality article layouts of various page sizes (left) and for complex articles with rich content (right).}
    \label{fig:hori_and_comp}
\end{figure}

\section{Evaluation} \label{sec:eval}

\begin{figure}[t]
    \centering
    \includegraphics[width=\linewidth]{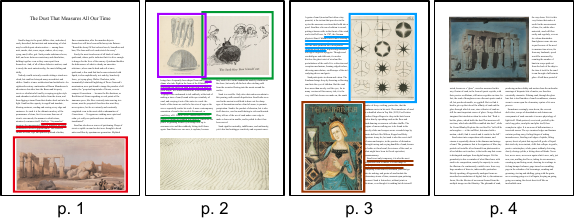}
    \caption{Our framework can produce high-quality multi-page article layouts without modifying the pipeline. Rectangles and underlines in the same colors indicate the text-illustration references.}
    \label{fig:multi}
\end{figure}

\begin{figure*}[t]
    \centering
    \includegraphics[width=\linewidth]{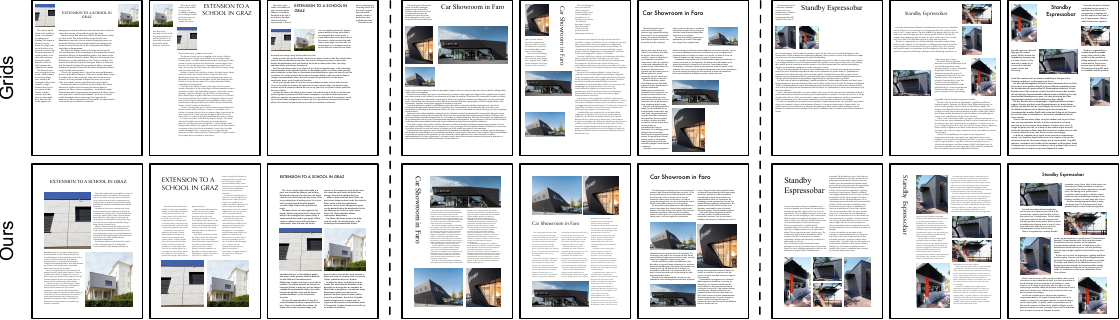}
    \caption{The article layouts created by GRIDS (top) and our method (bottom). Three articles are fed to both methods, and the corresponding results are separated by dashed lines. In each column, two article layouts created by different methods have the same number of headline, text, and illustration elements.}
    \label{fig:grids_comp}
\end{figure*}

\begin{figure*}[t]
    \centering
    \includegraphics[width=0.95\linewidth]{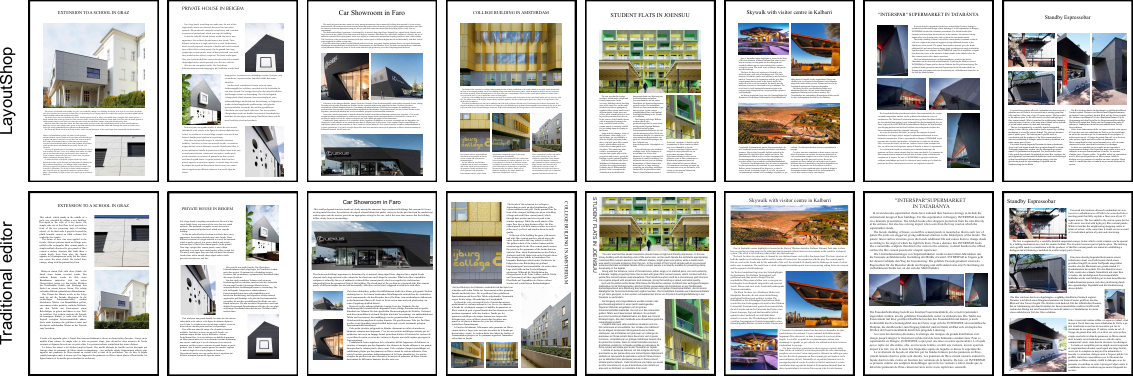}
    \caption{The representative article layouts created with \sysName\ and the traditional editor based on our provided 8 articles in Study I.}
    \label{fig:study1_examples}
\end{figure*}

We extensively evaluated our framework using articles from \emph{Architecture+Detail} and \emph{The Public Domain Review}.
Figures~\ref{fig:teaser}, \ref{fig:user_input}, \ref{fig:hori_and_comp}, \ref{fig:multi}, and \ref{fig:study1_examples} present representative results. While our tests focused on these sources, the framework is applicable to articles from other domains, since it uses only article content as geometric constraints.
Our framework allows users to create partial layout templates to specify design goals. By combining these with other templates, the system generates complete layouts tailored to user requirements (Figure~\ref{fig:user_input}).
The framework supports specialized designs (Figure~\ref{fig:special}). Users can overlay titles on background illustrations by specifying relationships or create irregular layouts by adding padding elements.
Layouts of various page sizes are supported (Figure~\ref{fig:hori_and_comp}, left) by adjusting $W$ and $H$ in Equation~\ref{eqn:page_size}.
With suitable templates, the framework efficiently generates high-quality layouts for complex, content-rich articles (Figure~\ref{fig:hori_and_comp}, right).
Multi-page layouts are also supported using the same pipeline (Figures~\ref{fig:teaser} and \ref{fig:multi}).
Additional results are provided in the supplemental materials.
We further evaluate the framework against a baseline (Section~\ref{sec:baseline}), providing produced layouts for qualitative comparison.
We also conducted a two-part user study. In Study I (Section~\ref{sec:study1}), participants created layouts using our framework or a mainstream editor to assess usability. In Study II (Section~\ref{sec:study2}), another group evaluated these layouts via an online questionnaire.

\begin{table*}[t]
\centering
\caption{Summary of the user study protocols for Study I (Section~\ref{sec:study1}) and Study II (Section~\ref{sec:study2}), detailing the participants, tasks, time budget, and evaluation metrics.}
\label{tab:user_study_protocol}
\begin{tabular}{lll}
\hline
\textbf{Protocol Aspect} & \textbf{Study I: Usability Study} & \textbf{Study II: Human Evaluation} \\ \hline
\textbf{Participants} & 12 ($6$ Novices, $6$ Designers) & 100 ($19$ with design background) \\
\textbf{Task} & Create 4 layouts per participant & Rate 48 generated layouts (randomized) \\
 & ($2$ via Our Framework vs. $2$ via Traditional) & on a 5-point Likert scale \\
\textbf{Data / Materials} & 8 real articles from \emph{Architecture+Detail} & 48 layouts created from Study I \\
\textbf{Time Budget} & $\sim$10 min tutorial + $\sim$30 min session & $\sim$10--15 min online questionnaire \\
\textbf{Metrics} & Completion time, Usability score (1-5), & Quality rating (1-5 Likert scale), \\
 & Network accuracy, Qualitative feedback & Statistical $t$-test comparisons \\ \hline
\end{tabular}%
\end{table*}

\subsection{Baseline comparison}\label{sec:baseline}
The method in \cite{jacobs2003adaptive} creates article layouts by adapting articles to existing templates, but cannot generate novel structures.
The method in \cite{zheng2019content} considers content but generates only layout templates. The method in \cite{ODonovan2014LearningLF} efficiently generates layouts for scattered elements (\eg, ads, fliers, posters) but does not address article layout constraints and therefore cannot produce tightly packed article layouts.
Although recent deep generative models can produce diverse bounding boxes, they cannot strictly enforce article constraints such as exact text fitting, image aspect ratios, and cross-page references.
Optimization-based methods therefore remain the most viable paradigm for guaranteeing these constraints, making GRIDS~\cite{dayama2020grids} the most feasible baseline despite its focus on GUI layouts.
However, its design objectives, \ie, overall alignment, rectangular outline, and element placement, are tailored to GUIs and are insufficient for articles. GRIDS also requires a fixed set of graphic elements, whereas article layouts may contain a variable number of text elements. To compare our method with GRIDS, we enumerate text elements from 2 to 4 as GRIDS input and add a text area constraint to ensure proper framing.

Figure~\ref{fig:grids_comp} compares article layouts generated by GRIDS and our method using the same articles.
Our method produces more reasonable and visually pleasing layouts by optimizing according to article layout design principles and constraining layout structures with user-preferred templates. In contrast, GRIDS searches a larger space defined by graphical elements and objectives without considering article design principles, often resulting in unreasonable structures. Our constrained search space enables higher-quality layouts with less computation: our method generates a layout in 106 ms, while GRIDS takes 1,974 ms. This demonstrates our method's suitability for article layout creation.

\subsection{Study I: usability study}\label{sec:study1}

This study investigates the usability of our framework (see Table~\ref{tab:user_study_protocol} for the complete user study protocol). As no comparable computational framework exists for creative article layouts, we compared our method with four mainstream editors: Adobe InDesign, Affinity Publisher, Canva, and Microsoft PowerPoint.
InDesign and Publisher offer powerful layout features but require more expertise, making them better suited to professionals. Canva and PowerPoint, with sufficient functions and simpler interfaces, are preferred by amateur users. Participants could freely choose their editor.
Participants created article layouts in two configurations, \ie, with our framework and with the traditional editor.
With our framework, participants first generated candidate layouts automatically, then refined them using interactive tools.
With the traditional editor, participants manually created article layouts, using features such as rulers, grid lines, and guides.

\paragraph*{Participants and apparatus.}
We recruited 6 university students (5 male, 1 female, aged 21-27) as novice users and 6 designers (1 male, 5 female, aged 23-36) as professionals. Two designers chose Adobe InDesign and Affinity Publisher, while the others used Canva and Microsoft PowerPoint, covering both professional publishing and consumer tools. While platforms like Figma are popular for UI, their core layout mechanisms overlap with these baseline editors.
The study was conducted on a MacBook Pro with an Apple M1 Pro processor and 16GB RAM.

\paragraph*{Task.}
The task involved freely creating article layouts using the provided articles. We supplied 8 real articles from \emph{Architecture+Detail}, an online magazine.
Each participant was assigned 4 articles and created 4 layouts: two with our framework and two with a traditional editor. The configuration order was counterbalanced across participants.
Participants did not create layouts for the same article twice to avoid bias. Article assignments ensured that each article was used equally and that configurations were balanced.
In total, 12 participants $\times$ 4 articles $=$ 48 article layouts were created in this study.

Before the study, participants received an introduction to our framework and practiced with an additional article. They also created a layout using the traditional editor for familiarity. The tutorial lasted about 10 minutes. During the study, participants created 4 layouts consecutively, with the entire session averaging 30 minutes per participant.

We also tested our layout assessment networks. When using our framework, participants saw both high-quality layouts (scores $>$ 0.5) and low-quality layouts (scores $<$ 0.5), with the layouts mixed and without indication of type. Their selections provided a measure of the network’s accuracy.

\paragraph*{Performance measures.}
We recorded the completion time for each layout creation and, when using our framework, the assessment scores of user-selected layouts.
Participants rated our framework’s usability and the quality of the produced layouts on a 5-point Likert scale and provided feedback.
Layouts from both configurations were saved for further evaluation in Study II.

\begin{figure*}
    \centering
    \includegraphics[width=\linewidth]{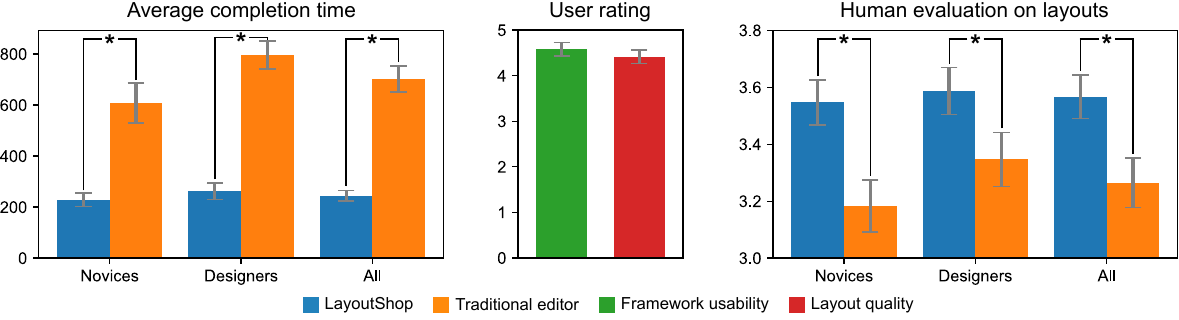}
    \caption{Left: the statistics of the average completion time for creating an article layout in the two configurations in Study I. Middle: the ratings on the usability of our framework and the qualities of the article layouts produced in Study I. Right: the statistics of the evaluation scores on the layouts created with our framework and with the traditional editor in Study II. The error bar represents the standard error of the mean. A $*$ indicates a significant difference between the data.}
    \label{fig:study_stat}
\end{figure*}

\paragraph*{Results.}
Figure~\ref{fig:study1_examples} shows representative layouts from both configurations using the 8 provided articles.
Figure~\ref{fig:study_stat} (left and middle) plots the statistics of the average completion time for creating an article layout in the two configurations, as well as the ratings on the usability of our framework and the qualities of its produced article layouts.
The participants created layouts more efficiently with our framework: average completion time was 245s (our framework) versus 702s (traditional editor).
A t-test confirmed a significant difference in completion time ($p<1.22\times10^{-9}$).
This trend held for both groups: designers averaged 261s (our framework) and 795s (traiditonal editor) ($p<6.86\times10^{-8}$); novices, 229s (our framework) and 609s (traiditonal editor) ($p<2.41\times10^{-4}$).
It is worth noting that the designers spent more time than novices, despite their greater skill, as they aimed for higher quality and spent more time selecting and editing templates. Novices generally made quicker selections.
Overall, the participants rated our framework highly for usability (4.6/5) and layout quality (4.4/5).

24 article layouts were created using our framework. We recorded the assessment scores provided by the networks when the participants selected these layouts. The average score was 0.91. 21 out of 24 scores were above 0.5. In our study, we deliberately mixed high- and low-scoring layouts in the list of candidate layouts. However, most of the selected layouts were high quality, \ie, with scores above 0.5. This indicates that the scores produced by our assessment networks align with human judgments of layout quality, thereby confirming their effectiveness.

We collected the participants' comments. Positive feedback highlights usability, efficiency, and support for novices, with users appreciating the diversity, quality, and resemblance of generated layouts to selected templates. Negative comments, mainly from designers, focused on limited editing freedom, lack of precise operations, and occasional unusual structures. Designers noted the inability to make specific modifications due to automatic layout optimization after each edit. 
In contrast, traditional editors allowed free editing but required time-consuming trial and error. 
Our framework prioritizes generating high-quality layouts, though at the expense of some editing flexibility. 
It is intended to suggest design ideas for users to finalize, rather than serving as an end-to-end solution.
Integrating our framework into traditional editors as a comprehensive tool would be promising.

\subsection{Study II: human evaluation}\label{sec:study2}

This study evaluated the layouts from Study I to determine whether our framework matches or surpasses manual design (Table~\ref{tab:user_study_protocol}).

\paragraph*{Participants.}
We recruited 100 participants (25 males, 75 females; aged 18–35), including 19 with graphic design training.

\paragraph*{Task.}
The participants rated layout quality using an online questionnaire.
All 48 layouts from 8 articles (created in Study I) were shown in random order. Participants rated each on a 5-point Likert scale, with 1 being strongly negative and 5 being strongly positive.

\paragraph*{Results.}
Figure~\ref{fig:study_stat} (right) shows average layout scores.
The 48 layouts were grouped by creation configuration: with our framework or with a traditional editor.
We computed two scores for each participant based on their ratings, corresponding to the two groups. 
On average, layouts from our framework scored higher: 3.57 (our framework) \vs 3.27 (traditional editor).
A paired t-test confirmed a significant difference ($p<5.16\times10^{-3}$) between the two groups.
To further examine the quality of article layouts, 
Dividing layouts by creator expertise led to the same result: designers’ layouts scored 3.59 (our framework) vs. 3.35 (traditional editor) ($p<2.90\times10^{-2}$); novices’ layouts scored 3.55 (our framework) \vs 3.18 (traditional editor) ($p<1.51\times10^{-3}$).
With traditional editors, designers outperformed novices (3.35 \vs 3.18), reflecting their greater expertise. With \sysName, the gap was negligible (3.59 \vs 3.55), as the network consistently produced high-quality layouts.
Our framework benefited both designers and novices, reducing effort and significantly improving layout quality.

\section{Conclusion}

We presented \sysName, a novel computational framework for exploratory design of creative article layouts. Inspired by actual layout design procedures, our algorithm lets users create or select layout templates to express preferences, constructs a design space, and extracts eligible layout structures. These structures frame article content through optimization. Two neural networks evaluate generated layouts, displaying only high-quality results for further user editing with interactive tools or existing editors.
Usability and evaluation studies confirm that our framework effectively assists users in creating high-quality article layouts.

\paragraph*{Limitations and future work.}
Although most results are high quality, some layouts produced by our framework may exhibit unusual structures because high-level design principles cannot be fully integrated into the optimization process.
Users, especially professionals, may find the interactive tools limiting, as some edits are constrained by the algorithm’s requirements for effective layouts.
While grid design is a standard approach for document layout~\cite{muller1996grid}, supporting non-axis-aligned designs is a promising future direction. It is also promising to integrate our framework into traditional editors as a comprehensive tool for such designs.

\section*{Acknowledgments}
We thank the anonymous reviewers for their constructive feedback and the user study participants for their time and effort. This work was supported by the National Natural Science Foundation of China (62472287) and the Natural Science Foundation of Shenzhen City (JCYJ20250604181519025).

\bibliographystyle{eg-alpha-doi} 
\bibliography{layoutshop}

@article{zheng2019content,
  title={Content-aware generative modeling of graphic design layouts},
  author={Zheng, Xinru and Qiao, Xiaotian and Cao, Ying and Lau, Rynson WH},
  journal={ACM TOG},
  volume={38},
  number={4},
  pages={1--15},
  year={2019},
  publisher={ACM New York, NY, USA}
}

@inproceedings{gupta2021layouttransformer,
  title={Layouttransformer: Layout generation and completion with self-attention},
  author={Gupta, Kamal and Lazarow, Justin and Achille, Alessandro and Davis, Larry S and Mahadevan, Vijay and Shrivastava, Abhinav},
  booktitle={ICCV},
  pages={1004--1014},
  year={2021}
}

@inproceedings{arroyo2021variational,
  title={Variational transformer networks for layout generation},
  author={Arroyo, Diego Martin and Postels, Janis and Tombari, Federico},
  booktitle={CVPR},
  pages={13642--13652},
  year={2021}
}

@inproceedings{tabata2019automatic,
  title={Automatic layout generation for graphical design magazines},
  author={Tabata, Sou and Yoshihara, Hiroki and Maeda, Haruka and Yokoyama, Kei},
  booktitle={ACM SIGGRAPH Posters},
  pages={1--2},
  year={2019}
}

@article{yang2016automatic,
  title={Automatic generation of visual-textual presentation layout},
  author={Yang, Xuyong and Mei, Tao and Xu, Ying-Qing and Rui, Yong and Li, Shipeng},
  journal={ACM TOMM},
  volume={12},
  number={2},
  pages={1--22},
  year={2016},
  publisher={ACM New York, NY, USA}
}

@inproceedings{lee2020guicomp,
  title={GUIComp: A GUI design assistant with real-time, multi-faceted feedback},
  author={Lee, Chunggi and Kim, Sanghoon and Han, Dongyun and Yang, Hongjun and Park, Young-Woo and Kwon, Bum Chul and Ko, Sungahn},
  booktitle={CHI},
  pages={1--13},
  year={2020}
}

@inproceedings{guo2021vinci,
  title={Vinci: an intelligent graphic design system for generating advertising posters},
  author={Guo, Shunan and Jin, Zhuochen and Sun, Fuling and Li, Jingwen and Li, Zhaorui and Shi, Yang and Cao, Nan},
  booktitle={CHI},
  pages={1--17},
  year={2021}
}

@inproceedings{zhang2020smarttext,
  title={Smarttext: Learning to generate harmonious textual layout over natural image},
  author={Zhang, Peiying and Li, Chenhui and Wang, Changbo},
  booktitle={ICME},
  pages={1--6},
  year={2020},
}

@inproceedings{vaddamanu2022harmonized,
  title={Harmonized Banner Creation from Multimodal Design Assets},
  author={Vaddamanu, Praneetha and Aggarwal, Vinay and Guda, Bhanu Prakash Reddy and Srinivasan, Balaji Vasan and Chhaya, Niyati},
  booktitle={CHI EA},
  pages={1--7},
  year={2022}
}

@book{lupton2014thinking,
  title={Thinking with type: A critical guide for designers, writers, editors, \& students},
  author={Lupton, Ellen},
  year={2014},
}

@book{muller1996grid,
  title={Grid systems in graphic design: A visual communication manual for graphic designers, typographers and three dimensional designers},
  author={M{\"u}ller-Brockmann, Josef},
  year={1996},
}

@inproceedings{dayama2020grids,
  title={Grids: Interactive layout design with integer programming},
  author={Dayama, Niraj Ramesh and Todi, Kashyap and Saarelainen, Taru and Oulasvirta, Antti},
  booktitle={CHI},
  pages={1--13},
  year={2020}
}

@inproceedings{swearngin2020scout,
  title={Scout: Rapid exploration of interface layout alternatives through high-level design constraints},
  author={Swearngin, Amanda and Wang, Chenglong and Oleson, Alannah and Fogarty, James and Ko, Amy J},
  booktitle={CHI},
  pages={1--13},
  year={2020}
}

@article{xu2022hierarchical,
  title={Hierarchical Layout Blending with Recursive Optimal Correspondence},
  author={Xu, Pengfei and Li, Yifan and Yang, Zhijin and Shi, Weiran and Fu, Hongbo and Huang, Hui},
  journal={ACM TOG},
  volume={41},
  number={6},
  pages={1--15},
  year={2022},
  publisher={ACM New York, NY, USA}
}

@inproceedings{kikuchi2021constrained,
  title={Constrained graphic layout generation via latent optimization},
  author={Kikuchi, Kotaro and Simo-Serra, Edgar and Otani, Mayu and Yamaguchi, Kota},
  booktitle={ACM MM},
  pages={88--96},
  year={2021}
}

@inproceedings{jyothi2019layoutvae,
  title={Layoutvae: Stochastic scene layout generation from a label set},
  author={Jyothi, Akash Abdu and Durand, Thibaut and He, Jiawei and Sigal, Leonid and Mori, Greg},
  booktitle={ICCV},
  pages={9895--9904},
  year={2019}
}

@inproceedings{jiang2022coarse,
author = {Jiang, Zhaoyun and Sun, Shizhao and Zhu, Jihua and Lou, Jian-Guang and Zhang, Dongmei},
title = {Coarse-to-Fine Generative Modeling for Graphic Layouts},
booktitle = {AAAI},
year = {2022},
month = {February},
}

@inproceedings{kong2022blt,
  title={BLT: bidirectional layout transformer for controllable layout generation},
  author={Kong, Xiang and Jiang, Lu and Chang, Huiwen and Zhang, Han and Hao, Yuan and Gong, Haifeng and Essa, Irfan},
  booktitle={ECCV},
  pages={474--490},
  year={2022},
}

@inproceedings{chai2023layoutdm,
  title={LayoutDM: Transformer-based Diffusion Model for Layout Generation},
  author={Chai, Shang and Zhuang, Liansheng and Yan, Fengying},
  booktitle={CVPR},
  pages={18349--18358},
  year={2023}
}

@inproceedings{inoue2023layoutdm,
    title={{LayoutDM: Discrete Diffusion Model for Controllable Layout Generation}},
    author={Naoto Inoue and Kotaro Kikuchi and Edgar Simo-Serra and Mayu Otani and Kota Yamaguchi},
    booktitle={CVPR},
    year={2023},
    pages={10167-10176},
  }

@inproceedings{hui2023unifying,
  title={Unifying Layout Generation with a Decoupled Diffusion Model},
  author={Hui, Mude and Zhang, Zhizheng and Zhang, Xiaoyi and Xie, Wenxuan and Wang, Yuwang and Lu, Yan},
  booktitle={CVPR},
  pages={1942--1951},
  year={2023}
}

@inproceedings{o2015designscape,
  title={Designscape: Design with interactive layout suggestions},
  author={Peter {O'Donovan} and Aseem Agarwala and Aaron Hertzmann},
  booktitle={CHI},
  pages={1221--1224},
  year={2015}
}

@inproceedings{Deka2017RicoAM,
  title={Rico: A Mobile App Dataset for Building Data-Driven Design Applications},
  author={Biplab Deka and Zifeng Huang and Chad Franzen and Joshua Hibschman and Daniel Afergan and Y. Li and Jeffrey Nichols and Ranjitha Kumar},
  booktitle={UIST},
  year={2017},
}

@inproceedings{Zhong2019PubLayNetLD,
  title={PubLayNet: Largest Dataset Ever for Document Layout Analysis},
  author={Xu Zhong and Jianbin Tang and Antonio Jimeno-Yepes},
  booktitle={ICDAR},
  year={2019},
  pages={1015-1022},
}

@article{Wang2018DeepCP,
  title={Deep convolutional priors for indoor scene synthesis},
  author={Kai Wang and Manolis Savva and Angel X. Chang and Daniel Ritchie},
  journal={ACM TOG},
  year={2018},
  volume={37},
  pages={1 - 14},
}

@inproceedings{Ritchie2018FastAF,
  title={Fast and Flexible Indoor Scene Synthesis via Deep Convolutional Generative Models},
  author={Daniel Ritchie and Kai Wang and Yu-An Lin},
  booktitle={CVPR},
  year={2018},
  pages={6175-6183},
}

@article{Wu2019DatadrivenIP,
  title={Data-driven interior plan generation for residential buildings},
  author={Wenming Wu and Xiaoming Fu and Rui Tang and Yuhan Wang and Yuanhang Qi and Ligang Liu},
  journal={ACM TOG},
  year={2019},
  volume={38},
  pages={1 - 12},
}

@inproceedings{Lee2019NeuralDN,
  title={Neural Design Network: Graphic Layout Generation with Constraints},
  author={Hsin-Ying Lee and Lu Jiang and Irfan Essa and Phuong B. Le and Haifeng Gong and Ming-Hsuan Yang and Weilong Yang},
  booktitle={ECCV},
  pages={491--506},
  year={2020},
}

@inproceedings{Patil2019READRA,
  title={READ: Recursive Autoencoders for Document Layout Generation},
  author={Akshay Gadi Patil and Omri Ben-Eliezer and Or Perel and Hadar Averbuch-Elor},
  booktitle={CVPR},
  year={2019},
  pages={2316-2325},
}

@inproceedings{Wang2020SceneFormerIS,
  title={SceneFormer: Indoor Scene Generation with Transformers},
  author={Xinpeng Wang and Chandan Yeshwanth and Matthias Nie{\ss}ner},
  booktitle={3DV},
  year={2020},
  pages={106-115},
}

@inproceedings{Paschalidou2021ATISSAT,
  title={ATISS: Autoregressive Transformers for Indoor Scene Synthesis},
  author={Despoina Paschalidou and Amlan Kar and Maria Shugrina and Karsten Kreis and Andreas Geiger and Sanja Fidler},
  booktitle={NeurIPS},
 year={2021},
}

@inproceedings{Jiang2022LayoutFormerCG,
  title={LayoutFormer++: Conditional Graphic Layout Generation via Constraint Serialization and Decoding Space Restriction},
  author={Zhaoyun Jiang and Jiaqi Guo and Shizhao Sun and Huayu Deng and Zhongkai Wu and V. Mijovi{\'c} and Zijiang James Yang and Jian-Guang Lou and D. Zhang},
  booktitle={CVPR},
  year={2022},
  pages={18403-18412},
}

@inproceedings{Li2019LayoutGANGG,
  title={LayoutGAN: Generating Graphic Layouts with Wireframe Discriminators},
  author={Jianan Li and Jimei Yang and Aaron Hertzmann and Jianming Zhang and Tingfa Xu},
  booktitle={ICLR},
  year={2019},
}

@article{Li2018GRAINS,
  title={GRAINS},
  author={Manyi Li and Akshay Gadi Patil and Kai Xu and Siddhartha Chaudhuri and Owais Khan and Ariel Shamir and Changhe Tu and Baoquan Chen and Daniel Cohen-Or and Hao Zhang},
  journal={ACM TOG},
  year={2018},
  volume={38},
  pages={1 - 16},
}

@article{xu2019global,
  title={Global beautification of 2D and 3D layouts with interactive ambiguity resolution},
  author={Xu, Pengfei and Yan, Guohang and Fu, Hongbo and Igarashi, Takeo and Tai, Chiew-Lan and Huang, Hui},
  journal={IEEE TVCG},
  volume={27},
  number={4},
  pages={2355--2368},
  year={2019},
  publisher={IEEE}
}

@misc{gurobi,
  author = {{Gurobi Optimization, LLC}},
  title = {{Gurobi Optimizer Reference Manual}},
  year = 2023,
  url = "https://www.gurobi.com"
}

@inproceedings{qiang2016learning,
  title={Learning to generate posters of scientific papers},
  author={Qiang, Yuting and Fu, Yanwei and Guo, Yanwen and Zhou, Zhi-Hua and Sigal, Leonid},
  booktitle={AAAI},
  volume={30},
  number={1},
  year={2016}
}

@article{tong2020directed,
  title={Directed graph convolutional network},
  author={Tong, Zekun and Liang, Yuxuan and Sun, Changsheng and Rosenblum, David S and Lim, Andrew},
  journal={arXiv preprint arXiv:2004.13970},
  year={2020}
}

@article{zhao2018characterizes,
  title={What characterizes personalities of graphic designs?},
  author={Zhao, Nanxuan and Cao, Ying and Lau, Rynson WH},
  journal={ACM TOG},
  volume={37},
  number={4},
  pages={1--15},
  year={2018},
  publisher={ACM New York, NY, USA}
}

@inproceedings{Todi2016SketchploreSA,
  title={Sketchplore: Sketch and Explore with a Layout Optimiser},
  author={Kashyap Todi and Daryl Weir and Antti Oulasvirta},
  booktitle={DIS},
  year={2016},
}

@inproceedings{Dayama2021InteractiveLT,
  title={Interactive Layout Transfer},
  author={Niraj Ramesh Dayama and Simo Santala and Lukas Br{\"u}ckner and Kashyap Todi and Jingzhou Du and Antti Oulasvirta},
  booktitle={UIST},
  year={2021},
}

@article{ODonovan2014LearningLF,
  title={Learning Layouts for Single-PageGraphic Designs},
  author={Peter O'Donovan and Aseem Agarwala and Aaron Hertzmann},
  journal={IEEE TVCG},
  year={2014},
  volume={20},
  pages={1200-1213},
}

@inproceedings{Jiang2020ORCSolverAE,
  title={ORCSolver: An Efficient Solver for Adaptive GUI Layout with OR-Constraints},
  author={Yue Jiang and Wolfgang Stuerzlinger and Matthias Zwicker and Christof Lutteroth},
  booktitle={CHI},
  year={2020},
}

@inproceedings{Duan2020OptimizingUI,
  title={Optimizing User Interface Layouts via Gradient Descent},
  author={Peitong Duan and Casimir Wierzynski and Lama Nachman},
  booktitle={CHI},
  year={2020},
}

@article{Kikuchi2021ModelingVC,
  title={Modeling Visual Containment for Web Page Layout Optimization},
  author={Kenji Kikuchi and Mayu Otani and Koh'ichiro Yamaguchi and Edgar Simo-Serra},
  journal={CGF},
  year={2021},
  volume={40},
}

@inproceedings{DameraVenkata2011ProbabilisticDM,
  title={Probabilistic document model for automated document composition},
  author={Niranjan Damera-Venkata and Jos{\'e} Bento and Eamonn O'Brien-Strain},
  booktitle={DocEng},
  year={2011},
}

@inproceedings{Gross1996AmbiguousIA,
  title={Ambiguous intentions: a paper-like interface for creative design},
  author={Mark D. Gross and Ellen Yi-Luen Do},
  booktitle={UIST},
  year={1996},
}

@inproceedings{Herring2009GettingIU,
  title={Getting inspired!: understanding how and why examples are used in creative design practice},
  author={Scarlett R. Herring and Chia-Chen Chang and Jesse Krantzler and Brian P. Bailey},
  booktitle={CHI},
  year={2009},
}

@inproceedings{Hsu2023PosterLayoutAN,
  title={PosterLayout: A New Benchmark and Approach for Content-Aware Visual-Textual Presentation Layout},
  author={Hsiao-An Hsu and Xiangteng He and Yuxin Peng and Hao-Song Kong and Qing Zhang},
  booktitle={CVPR},
  year={2023},
  pages={6018-6026},
}

@inproceedings{Zhou2022CompositionawareGL,
  title={Composition-aware Graphic Layout GAN for Visual-textual Presentation Designs},
  author={Min Zhou and Chenchen Xu and Ye Ma and Tiezheng Ge and Yuning Jiang and Weiwei Xu},
  booktitle={IJCAI},
  pages={4995--5001},
  year={2022},
}

@inproceedings{Cao2022GeometryAV,
  title={Geometry Aligned Variational Transformer for Image-conditioned Layout Generation},
  author={Yunning Cao and Ye Ma and Min Zhou and Chuanbin Liu and Hongtao Xie and Tiezheng Ge and Yuning Jiang},
  booktitle={ACM MM},
  year={2022},
}

@article{jacobs2003adaptive,
  title={Adaptive grid-based document layout},
  author={Jacobs, Charles and Li, Wilmot and Schrier, Evan and Bargeron, David and Salesin, David},
  journal={ACM TOG},
  volume={22},
  number={3},
  pages={838--847},
  year={2003},
  publisher={ACM New York, NY, USA}
}

@inproceedings{schrier2008adaptive,
  title={Adaptive layout for dynamically aggregated documents},
  author={Schrier, Evan and Dontcheva, Mira and Jacobs, Charles and Wade, Geraldine and Salesin, David},
  booktitle={IUI},
  pages={99--108},
  year={2008}
}

@inproceedings{kumar2011bricolage,
  title={Bricolage: example-based retargeting for web design},
  author={Kumar, Ranjitha and Talton, Jerry O and Ahmad, Salman and Klemmer, Scott R},
  booktitle={CHI},
  pages={2197--2206},
  year={2011}
}

@inproceedings{zheng2019faceoff,
  title={Faceoff: Assisting the manifestation design of web graphical user interface},
  author={Zheng, Shuyu and Hu, Ziniu and Ma, Yun},
  booktitle={WSDM},
  pages={774--777},
  year={2019}
}

@article{laine2021responsive,
  title={Responsive and personalized web layouts with integer programming},
  author={Laine, Markku and Zhang, Yu and Santala, Simo and Jokinen, Jussi PP and Oulasvirta, Antti},
  journal={PACM HCI},
  volume={5},
  number={EICS},
  pages={1--23},
  year={2021},
  publisher={ACM New York, NY, USA}
}

@inproceedings{chen2023docdancer,
  title={DocDancer: Authoring Ultra-responsive Documents with Layout Generation},
  author={Chen, Yuexi and Liu, Zhicheng and Tensmeyer, Christopher and Elmqvist, Niklas and Morariu, Vlad I},
  booktitle={VL/HCC},
  pages={133--138},
  year={2023},
}

@inproceedings{he2023diffusion,
  title={Diffusion-based document layout generation},
  author={He, Liu and Lu, Yijuan and Corring, John and Florencio, Dinei and Zhang, Cha},
  booktitle={ICDAR},
  pages={361--378},
  year={2023},
}

@inproceedings{levi2023dlt,
  title={Dlt: Conditioned layout generation with joint discrete-continuous diffusion layout transformer},
  author={Levi, Elad and Brosh, Eli and Mykhailych, Mykola and Perez, Meir},
  booktitle={ICCV},
  pages={2106--2115},
  year={2023}
}

@inproceedings{chen2024towards,
  title={Towards aligned layout generation via diffusion model with aesthetic constraints},
  author={Chen, Jian and Zhang, Ruiyi and Zhou, Yufan and Chen, Changyou},
  booktitle={ICLR},
  year={2024}
}

@inproceedings{lin2023parse,
  title={A parse-then-place approach for generating graphic layouts from textual descriptions},
  author={Lin, Jiawei and Guo, Jiaqi and Sun, Shizhao and Xu, Weijiang and Liu, Ting and Lou, Jian-Guang and Zhang, Dongmei},
  booktitle={ICCV},
  pages={23622--23631},
  year={2023}
}

@inproceedings{lin2023layoutprompter,
  title={Layoutprompter: awaken the design ability of large language models},
  author={Lin, Jiawei and Guo, Jiaqi and Sun, Shizhao and Yang, Zijiang and Lou, Jian-Guang and Zhang, Dongmei},
  booktitle={NeurIPS},
  volume={36},
  pages={43852--43879},
  year={2023}
}

@inproceedings{inoue2023towards,
  title={Towards flexible multi-modal document models},
  author={Inoue, Naoto and Kikuchi, Kotaro and Simo-Serra, Edgar and Otani, Mayu and Yamaguchi, Kota},
  booktitle={CVPR},
  pages={14287--14296},
  year={2023}
}

@inproceedings{shabani2024visual,
  title={Visual layout composer: Image-vector dual diffusion model for design layout generation},
  author={Shabani, Mohammad Amin and Wang, Zhaowen and Liu, Difan and Zhao, Nanxuan and Yang, Jimei and Furukawa, Yasutaka},
  booktitle={CVPR},
  pages={9222--9231},
  year={2024}
}

@inproceedings{horita2024retrieval,
  title={Retrieval-augmented layout transformer for content-aware layout generation},
  author={Horita, Daichi and Inoue, Naoto and Kikuchi, Kotaro and Yamaguchi, Kota and Aizawa, Kiyoharu},
  booktitle={CVPR},
  pages={67--76},
  year={2024}
}

@inproceedings{yu2024layoutdetr,
  title={Layoutdetr: detection transformer is a good multimodal layout designer},
  author={Yu, Ning and Chen, Chia-Chih and Chen, Zeyuan and Meng, Rui and Wu, Gang and Josel, Paul and Niebles, Juan Carlos and Xiong, Caiming and Xu, Ran},
  booktitle={ECCV},
  pages={169--187},
  year={2024},
}

@inproceedings{seol2024posterllama,
  title={PosterLlama: Bridging Design Ability of Language Model to Content-Aware Layout Generation},
  author={Seol, Jaejung and Kim, Seojun and Yoo, Jaejun},
  booktitle={ECCV},
  pages={451--468},
  year={2024},
}

@article{shiripour2021grid,
  title={Grid-based genetic operators for graphical layout generation},
  author={Shiripour, Morteza and Dayama, Niraj Ramesh and Oulasvirta, Antti},
  journal={PACM HCI},
  volume={5},
  number={EICS},
  pages={1--30},
  year={2021},
  publisher={ACM New York, NY, USA}
}

@article{hu2025structlayoutformer,
  title={StructLayoutFormer: Conditional Structured Layout Generation via Structure Serialization and Disentanglement},
  author={Hu, Xin and Xu, Pengfei and Zhou, Jin and Fu, Hongbo and Huang, Hui},
  journal={IEEE TVCG},
  year={2025},
  publisher={IEEE}
}

@article{xu2026gtlayout,
  author={Xu, Pengfei and Shi, Weiran and Hu, Xin and Fu, Hongbo and Huang, Hui},
  journal={Computational Visual Media}, 
  title={GTLayout: Learning General Trees for Structured Grid Layout Generation}, 
  year={2026},
  volume={12},
  number={3},
  pages={677-699},
}

@inproceedings{chai2023two,
  title={Two-stage content-aware layout generation for poster designs},
  author={Chai, Shang and Zhuang, Liansheng and Yan, Fengying and Zhou, Zihan},
  booktitle={ACM MM},
  pages={8415--8423},
  year={2023}
}

@inproceedings{jiang2025illustration,
  title={Illustration Layout Generation for Slide Enhancement with Pixel-based Diffusion Model},
  author={Jiang, Zhaoyun and Guo, Jiaqi and Liu, Shakie and Han, Chao and Liu, Ting and Lou, Jian-Guang and Zhang, Dongmei},
  booktitle={ACM MM},
  pages={5365--5374},
  year={2025}
}

@article{guo2025contentdm,
  title={ContentDM: A Layout Diffusion Model for Content-Aware Layout Generation},
  author={Guo, Honglin and Nie, Weizhi and Chen, Ruidong and Wang, Lanjun and Jin, Guoqing and Liu, Anan},
  journal={IEEE TAI},
  year={2025},
  publisher={IEEE}
}

@inproceedings{tang2024layoutkag,
  title={LayoutKAG: Enhancing Layout Generation in Large Language Models Through Knowledge-Augmented Generation},
  author={Tang, Hongbo and Zhao, Shuai and Luo, Jing and Su, Yihang and Yang, Jinjian},
  booktitle={AIHCIR},
  pages={292--299},
  year={2024},
}

@inproceedings{zhang2024vascar,
  title={Vascar: Content-aware layout generation via visual-aware self-correction},
  author={Zhang, Jiahao and Yoshihashi, Ryota and Kitada, Shunsuke and Osanai, Atsuki and Nakashima, Yuta},
  booktitle={MIRU},
  year={2025}
}

@article{zhang2025smaller,
  title={Smaller But Better: Unifying Layout Generation with Smaller Large Language Models},
  author={Zhang, Peirong and Zhang, Jiaxin and Cao, Jiahuan and Li, Hongliang and Jin, Lianwen},
  journal={IJCV},
  volume={133},
  number={7},
  pages={3891--3917},
  year={2025},
  publisher={Springer}
}

@inproceedings{cheng2025graphic,
  title={Graphic design with large multimodal model},
  author={Cheng, Yutao and Zhang, Zhao and Yang, Maoke and Nie, Hui and Li, Chunyuan and Wu, Xinglong and Shao, Jie},
  booktitle={AAAI},
  volume={39},
  number={3},
  pages={2473--2481},
  year={2025}
}

@inproceedings{lu2025uni,
  title={Uni-layout: Integrating human feedback in unified layout generation and evaluation},
  author={Lu, Shuo and Chen, Yanyin and Feng, Wei and Fan, Jiahao and Li, Fengheng and Zhang, Zheng and Lv, Jingjing and Shen, Junjie and Law, Ching and Liang, Jian},
  booktitle={ACM MM},
  pages={7709--7718},
  year={2025}
}

@INPROCEEDINGS {Hsu2025PosterO,
author = { Hsu, HsiaoYuan and Peng, Yuxin },
booktitle = {CVPR},
title = { PosterO: Structuring Layout Trees to Enable Language Models in Generalized Content-Aware Layout Generation },
year = {2025},
pages = {8117-8127},
}

@INPROCEEDINGS{Teng2025PosterCoT,
author = {Teng, Fei and Gao, Mengjiao and Wang, Long and Tan, Liwei and Liu, He and Li, Xiaoshu and Wang, Xuejian and Huang, Sitong and Zhang, Xiaolu},
title = {PosterCoT: Poster Layout Design Model Using Multi-Modal Training and Chain-of-Thought Enhancement},
year = {2025},
booktitle = {AICI},
pages = {52–57},
}

@inproceedings{Chen2025T-Stars-Poster,
author = {Chen, Hongyu and Zhou, Min and Jiang, Jing and Chen, Jiale and Lu, Yang and Lin, Zihang and Xiao, Bo and Ge, Tiezheng and Zheng, Bo},
title = {T-Stars-Poster: A Framework for Product-Centric Advertising Image Design},
year = {2025},
booktitle = {CIKM},
pages = {5583–5591},
}

@article{Yang2025OrderMatters,
author = {Yang, Bo and Cao, Ying},
title = {Order Matters: Learning Element Ordering for Graphic Design Generation},
year = {2025},
issue_date = {August 2025},
address = {New York, NY, USA},
volume = {44},
number = {4},
issn = {0730-0301},
journal = {ACM TOG},
month = jul,
articleno = {34},
numpages = {16},
}

@inproceedings{Kikuchi2025MarkupDM,
author = {Kikuchi, Kotaro and Honda, Ukyo and Inoue, Naoto and Otani, Mayu and Simo-Serra, Edgar and Yamaguchi, Kota},
title = {Multimodal Markup Document Models for Graphic Design Completion},
year = {2025},
booktitle = {ACM MM},
pages = {11022–11031},
}

@article{cao2012automatic,
  title={Automatic stylistic manga layout},
  author={Cao, Ying and Chan, Antoni B and Lau, Rynson WH},
  journal={ACM TOG},
  volume={31},
  number={6},
  pages={1--10},
  year={2012},
  publisher={ACM New York, NY, USA}
}

@article{yang2021automatic,
  title={Automatic comic generation with stylistic multi-page layouts and emotion-driven text balloon generation},
  author={Yang, Xin and Ma, Zongliang and Yu, Letian and Cao, Ying and Yin, Baocai and Wei, Xiaopeng and Zhang, Qiang and Lau, Rynson WH},
  journal={ACM TOMM},
  volume={17},
  number={2},
  pages={1--19},
  year={2021},
  publisher={ACM New York, NY, USA}
}

@article{chen2024manga,
  title={Manga generation via layout-controllable diffusion},
  author={Chen, Siyu and Li, Dengjie and Bao, Zenghao and Zhou, Yao and Tan, Lingfeng and Zhong, Yujie and Zhao, Zheng},
  journal={arXiv preprint arXiv:2412.19303},
  year={2024}
}


\end{document}